\documentclass[a4paper,12pt,review,twocolumn]{IEEEtran}
\newtheorem{thm}{Theorem}[section]

\newtheorem{lem}[thm]{Lemma}

\newtheorem{defn}[thm]{Definition}
\usepackage{graphicx}
\usepackage{amsmath,amssymb}

\begin{document}

\title{Generalized PMC model for the hybrid diagnosis of multiprocessor systems}

\author{Qiang Zhu
\IEEEcompsocitemizethanks{\IEEEcompsocthanksitem The author is with the School
of Mathematics and Statistics, Xidian University, Xi'an, Shaanxi, 710071, China.
E-mail: zhuustcer@gmail.com}}

\maketitle

\begin{abstract}
Fault diagnosis is important to the design and maintenance of large multiprocessor systems. PMC model is the most famous diagnosis model in the system level diagnosis of multiprocessor systems. Under the PMC model, only node faults are allowed. But in real circumstances, link faults may occur. So based on the PMC model, we propose in this paper a diagnosis model called the generalized PMC(GPMC) model to adapt to the real circumstances.  The foundation of GPMC model has been  established.  And to measure the fault diagnosis capability of multiprocessor systems under the GPMC model, the fault diagnosis capability measuring parameters:  $h$-edge restricted diagnosability and $h$-vertex restricted edge diagnosability have been introduced. As an application, the $h$-edge restricted diagnosability and $h$-vertex restricted edge diagnosability of hypercubes are explored.
\end{abstract}

\IEEEkeywords
{ \ Interconnection Networks, PMC model, Multiprocessor systems, Fault diagnosis}

\section{Introduction}
High performance computing is essential to the competitiveness of a country. Nowadays some super computers even have hundreds of thousands of processors. With so many computers and communication links, it's  inevitable that node or link faults may occur in the system. Thus
Fault diagnosis is important to the design and maintenance of large multiprocessor systems.
System level diagnosis is an approach for fault diagnosis of multiprocessor systems.   Different definitions of test and different assumptions on test result lead to different diagnosis models.

PMC model, proposed by F.P. Preparata, G. Metze, \&  R. Chien, is the most famous and most widely studied model in system level diagnosis\cite{F.P.Preparata1967}. Under the PMC model, it is assumed that only node faults can occur. But in real circumstances, both node and link faults may occur. So it's significant to study the fault diagnosis of multiprocessor systems under hybrid fault circumstances. Based on the PMC model, we have introduced the generalized PMC (GPMC) model to adapt to  the  hybrid fault circumstances. 

A multiprocessor system can be modeled with a graph called its interconnection network when each node is represented by a vertex and each communication link represented by an edge. In the following, a multiprocessor system and its interconnection network are used interchangeably. The performance of a multiprocessor system is greatly influenced by its interconnection network. So various interconnection networks have been proposed. In the design and maintenance of a multiprocessor system, it's necessary to explore the properties of its interconnection network.  Among all the properties, fault diagnosis capability of an interconnection network is quite important to evaluate its suitability as the underlying topology of a fault tolerant high performance computing system. Hypercubes is one of the most famous interconnection network. The second contribution of this paper is the study of the fault diagnosis capability of hypercubes under the GPMC model. The methods can be used to evaluate the diagnosis capability of other interconnection networks.

The following of this paper is organized as follows: Notations and Preliminaries  used in this paper are presented in section 2;  In section 3,  first the basic of PMC models are presented, then  the foundations of generalized PMC model are established; In section 4, we study the fault diagnosis capability of hypercubes under the generalized PMC model. Section 5 concludes the paper and present some advice for possible future works.
\section{Notations and Terminologies}
 For notations and terminologies not defined here, we follow \cite{J.A.Bondy1976}. Given a simple undirected graph $G=(V, E)$, $V(G)$ and $E(G)$ are used to denote the vertex set and edge set of  $G$ respectively.  For an edge $e$ with $u,v$ as its end-vertices, we use $uv$ to denote $e$ and call $u, v$ are adjacent to each other and $e$ is incident to both $u$ and $v$. Given a vertex $u$ in $G$, we use $N(u)$(resp.  $NE(u)$ ) to denote the set of all its adjacent vertices (resp. incident edges) in $G$.   $d(u)$ is defined to be the number of vertices in $N(u)$, called the degree of $u$. $\delta(G)$ is defined to be the minimum degree over all the vertices in $G$.

\section{PMC model and Generalized PMC model}
\subsection{Basics of the PMC model}
System level diagnosis is an important approach for the fault diagnosis of multiprocessor systems. In this approach, diagnosis is performed by mutual tests of processors in the system. The set of all  test results is called a syndrome of the system. Then based on the assumptions on the test results,  the faulty processors are located according to the syndrome of the system. Different definitions of test and different assumptions on test results lead to different diagnosis models.

Proposed by  Preparata, Metze and Chien\cite{F.P.Preparata1967}, PMC model is the most famous and most widely studied model in the system level diagnosis of multiprocessor systems. Under the PMC model, it is assumed that there is no link faults and only adjacent processors can test the status of each other. All node faults are permanent, and a node fault can always be detected by a fault-free vertex. Under the PMC model, a test can be represented by an ordered pair $(u, v)$ where $u$ is the tester and $v$ is the testee. The result of the test $(u,v)$ is denoted by $r(u, v)$. It is 0 if $u$ evaluates $v$ as fault-free and $1$ if $u$ evaluates $v$ as faulty. Under the PMC model, it is assumed that test result $r(u,v)$ is reliable if and only if  the tester $u$ is fault-free. That is, if $u$ is fault-free, then $r(u,v)=0$ means that $v$ is fault-free and $r(u,v)=1$ means that $v$ is faulty; If the tester $u$ is faulty, then the status of $v$ is irrelevant
 to  the test result $r(u,v)$.

Given a multiprocessor system, the set of all test results is called a syndrome of the multiprocessor system. Under the PMC model, a fault set $F$ is said to be consistent with a syndrome $\sigma$  if $\sigma$ can be aroused by the circumstance that all nodes in $F$ are faulty and all nodes not in $F$ are fault-free.
 Since the test result of a faulty tester is unreliable, a fault set $F$ can be consistent with many syndromes, the set of all syndromes consistent with $F$ is denoted by $\sigma(F)$.
Two faulty sets $F_1, F_2$ are distinguishable if and only if $\sigma(F_1)\cap \sigma(F_2)=\emptyset$. Otherwise, they are indistinguishable. Since the test result of a faulty vertex is unreliable, $V(G)$ is consistent with any syndrome of $G$. Thus to locate faulty vertices, people often suppose there exists an upper bounder for the number of faulty vertices.
A multiprocessor system $G$ is $t$-diagnosable if and only if all the faulty vertices can be guaranteed to be located provided that the number of faulty vertices does not exceed $t$.  The diagnosability of $G$ is the maximum integer $t$ such that $G$ is $t$-diagnosable. The diagnosability of a multiprocessor system can measure its fault diagnosis capability. The diagnosability of many interconnection networks have been explored\cite{Hsieh2008,Ahlswede2008,Stephens1993,Shin1993,
Huang1989,Krawczyk1988,Otsuka1990,Caruso2002,Krawczyk1988-2}.

 \subsection{The generalized PMC model }
To adapt to the  hybrid fault circumstances, we must make some modifications to the traditional PMC model.
\begin{itemize}
\item {\bf Definition of test}

Similar to the PMC model, under the generalized PMC model, it is  assumed that a test $t(u,v;e)$ involves two adjacent processors $u, v$ and the edge $e=uv$ between them. In the test $t(u,v;e)$, $u$ is called the tester, $v$ is called the tested vertex and $e$ is called the test edge.
\item {\bf Assumptions}

1. The links incident to any faulty processors are good.

The assumption is justified based on the following analysis: 1. When a link has a faulty end-vertex, it's of no use; 2. When a faulty processors is replaced or removed, all its incident links have to be rechecked or removed. So it's not necessary to determine their status. 3. When a link has a faulty end-vertex,it's not possible to determine its status under the GPMC model.

2. The test result of a good tester is reliable and test result of a faulty tester is unreliable.

That is, for a test $t(u, v; e)$, if $u$ is good, the test passed whenever  $v, e$ are both good. If $u$ is bad, the test result is irrelevant to the status of $v, e$.

This assumption is similar to the corresponding assumption in the PMC model. So it's justification is obvious.
\end{itemize}
Based on Assumption 1, we propose the definition of consistent faulty pairs in the hybrid fault circumstances.
\begin{defn}
Given a multiprocessor system $G$, we can $(F,S)$ a consistent faulty pair of $G$ if all vertices in $F$ cannot be incident to any edge in $S$.
 \end{defn}

  Similar to the PMC model, a syndrome in the generalized PMC model is the set of all  the test results. A syndrome $\sigma$ is said to be consistent to a faulty pair $(F,S)$ is the syndrome can be aroused in the circumstance that all vertices in $F$ and all edges in $S$ are faulty and all the other vertices and edges are faulty-free. Since the test result of a faulty tester is unreliable. The syndromes consistent to a faulty pair $(F,S)$ may not be unique. We use $\sigma(F,S)$ to denote the set of all syndromes consistent with $(F,S)$.
 Based on Assumption 1, given a syndrome $\sigma$ of a multiprocessor system $G$  under the GPMC model,  the purpose of diagnosis is to find a desired consistent faulty pair $(F,S)$ consistent with $\sigma$.

  \begin{defn}
  {\bf Distinguishable faulty pairs：} Two faulty pairs $(F_1, S_1)$ and $(F_2, S_2)$ are distinguishable if and only if $\sigma(F_1,S_1)\cap \sigma(F_2, S_2)=\emptyset$. Otherwise, they are indistinguishable.
\end{defn}
The diagnosis of a multiprocessor system is the process of locating faulty processors and faulty links according to a syndrome of the system. If all processors in a multiprocessor system are faulty, then any syndrome can be aroused. So under the generalized PMC model any syndrome is consistent with $(V(G), S)$ where $S$ is any edge subset of $G$.  For a syndrome $\sigma$, we use $FS(\sigma)$ to denote the set of all faulty pairs consistent with it.  In this circumstance,  how to judge which faulty pair is the desired faulty pair  is a challenging problem. Under the PMC model,  some researchers suppose that there exists an upper bound for the number of faulty vertices. A system is called $t$-diagnosable under the PMC model if all the faulty processors can be guaranteed to be located provided that the number of faulty processors does not exceed $t$. The diagnosability of a system $G$ is the biggest integer $t$ such that $G$ is $t$-diagnosable. Diagnosability is an important measure of the diagnosis capability of multiprocessor systems under the PMC model. Under the generalized PMC model, we propose the definition of $(t,s)$-diagnosable systems and $h$-restricted vertex diagnosability and $r$-restricted edge diagnosability to measure the diagnosis capability of interconnection networks.
\begin{defn}
{\bf $(t,s)$-diagnosable }Let $t, s$ be two natural numbers, a multiprocessor system $G$ is $(t,s)$ diagnosable if and only any two distinct faulty pairs $(F_1, S_1)$ and $(F_2, S_2)$ with $|F_1|,|F_2|\le t$ and $|S_1|,|S_2|\le s$ are distinguishable.
\end{defn}
\begin{defn}
Given a multiprocessor system $G$ and two integers $h,r$, the $h$-restricted vertex diagnosability of $G$ is the maximum integer $t$ such that $G$ is $(t,h)$-diagnosable, denoted by $t_h^{e}(G)$; Similarly, the $r$-restricted edge diagnosability of $G$ is the maximum integer $s$ such that $G$ is $(r, s)$-diagnosable, denoted by $s_r^{v}(G)$.
\end{defn}

The $h$-restricted vertex diagnosability and $r$-restricted edge diagnosability under the generalized PMC model  can be viewed as a generalization of the diagnosability under the PMC model, and thus can reflect the fault diagnosis capability of interconnection networks under the GPMC model.

To locate faulty processors and faulty links, we need to characterize distinguishable faulty pairs. The following lemma characterizes two distinguishable faulty pairs.
\begin{lem}\label{dis}
Given a multiprocessor systems $G(V,E)$, $F_1, F_2\subset V(G)$ and $S_1, S_2\subset E(G)$. Two distinct consistent faulty pairs $(F_1, S_1)$, $(F_2, S_2)$ are distinguishable under the generalized PMC model if and only if at least one of the following two conditions is satisfied.
\begin{itemize}
\item 1) There exists an edge $e=uv$ such that $u\in F_1 -F_2$(resp. $u\in F_2 -F_1$ ), $v\in V-F_1\cup F_2$ such that the edge $e\not\in S_2$(resp. $S_1$).

\item 2) There exists an edge $e=uv\in S_1 -S_2$(resp. $e=uv\in S_2 -S_1$) such that $u, v\not \in F_2$(resp. $u, v\not \in F_1$).
\end{itemize}
\end{lem}
{\textbf Proof: } 1) The sufficiency of the two conditions are  easy to verify. For both conditions,  in the former case(resp.  in the latter case ) the test result  $r(v,u;e)=1$ (resp.  $r(v,u;e)=0$) in the fault circumstance $(F_1, S_1 )$ (resp. $(F_2, S_2 )$). Thus no syndrome can be aroused in both fault circumstances,  $(F_1, S_1 )$ and $(F_2, S_2 )$ are distinguishable.

2) The necessity of the conditions is presented in this paragraph. Since  $(F_1, S_1 )$ and $(F_2, S_2 )$ are distinguishable, there must exist a test $t(u,v;e)$ whose result $r(v,u;e)$ must be different in the two fault circumstances.  Without loss of generality, suppose $r(v,u;e) $ must be 1 in the fault circumstance $(F_1,S_1)$ and $r(v,u;e) $ must be 0 in the fault circumstance $(F_2, S_2)$. Since $r(v,u;e) $ must be 1 in the fault circumstance $(F_1,S_1)$, $v\not\in F_1$ and either $u\in F_1$ or $e\in S_1$.  Since $r(v,u;e) $ must be 0 in the fault circumstance $(F_2, S_2)$, $u, v\not\in F_2$ and $e\not\in S_2$. That is,

Either
there exists an edge $e=uv$ such that $u\in F_1 -F_2$, $v\in V-F_1\cup F_2$ such that the edge $e\not\in S_2$.
Or there exists an edge $e=uv\in S_1 -S_2$ such that $u, v\not \in F_2$. \hfill\rule{1mm}{2mm}

In the following Lemma we present some basic results about the $s$-restricted diagnosability and $r$-restricted edge diagnosability of an interconnection network $G$.
\begin{lem}\label{diagboudary}
Given a multiprocessor $G$ with minimum degree $\delta(G)$ and $m$ edges, let   $t(G)$ be the diagnosability of $G$ under the PMC model, we have

1) $t_0(G)=t(G)$, $t_h(G)\le \delta(G)-h$ for $0\le h\le \delta(G)$.

2) $s_0(G)=m$,  $s_1(G)\le \delta(G)-2$.
\end{lem}
{\textbf Proof}  By the definition of  $t_s(G)$, it's obvious that $t_0(G)=t(G)$.
Let $u$ be a vertex of $G$ with $d(u)=\delta(G)$. Suppose  $N(u)=\{u^1, u^2, \cdots u^{\delta(G)}\}$, $NE(u)=\{e_1, e_2, \cdots e_{\delta(G)}\}$ where $e_i=(u, u^i)$. Let $F_1=\{u, u^{h+1},  u^{h+2}, \cdots u^{\delta(G)}\}$, $F_2=  \{ u^{h+1},  u^{h+2}, \cdots u^{\delta(G)}\}$, $S_1=\emptyset, S_2= \{e_1, e_2, \cdots e_h\}$. Then it's clear that both $(F_1, S_1)$ and $(F_2, S_2)$ are consistent faulty pairs. And by Lemma \ref{dis}, the two pairs are indistinguishable. Thus $t_h(G)\le \delta(G)-h$.

2) When there is no faulty vertices, the status of any faulty edge can be determined by the test result involving it, so $s_0(G)=m$;  Let $e=(u, v)$ where $d(u)=\delta(G)$. Suppose $NE(u)=\{e, e_1, e_2, \cdots e_{\delta(G)-1}\}$.  Let $F_1=\{u\}$, $F_2=\{v\}$, $S_1=\emptyset$, $S_2=\{e_1, e_2, \cdots e_{\delta(G)-1}\}. $ It's clear that $(F_1, S_1)$ and $(F_2, S_2)$ are both consistent faulty pairs and they are indistinguishable. So $s_1(G)\le \delta(G)-2$. \hfill\rule{1mm}{2mm}

The above lemma shows that when we explore the $h$-edge-restricted diagnosability of $G$, we only need to explore the case of $h\le \delta(G)$. And the $1$-vertex-restricted edge diagnosability of $G$ may  not equal its 1-edge-restricted diagnosability. We may need quite distinct methods to explore the $r$-vertex-restricted edge diagnosability of $G$.  There is much to explore.

\section{Fault diagnosis capability of hypercubes under the GPMC model}
In this section, we will explore the diagnosis capability of hypercubes under hybrid fault circumstances.
\subsection{Preliminaries of hypercubes}
Hypercubes are one of the most famous interconnection networks, due to its many attractive properties such as high symmetry, good graph embedding and etc.The properties of hypercubes have been extensively studied\cite{Kung2009-p801-811,Zhu2006-p111-121,J.-M.Xu2005-p203-207,Wang1995-p136-142,Yang2003//-p5-50,Wang1995//-p42-136}. Hypercubes have been used in the design of large multiprocessor systems\cite{Nugent1988-p51-60,R.Arlanskas1988-p38-42}.

Each vertex of an $n$-dimensional hypercube $Q_n$ can be labelled with an $n$-bit binary string, two vertices $u, v$ are adjacent if and only if their labels differ in exactly 1 bit position. Thus $Q_n$ has $2^n$ vertices and the degree of each vertex is $n$.
For a vertex $u$ in $Q_n$, we use $u^i$ to denote the neighbor of $u$ who differ with $u$ only in the $i$-th bit position. That is, $u^i=u_1\cdots u_{i-1}\overline{u_i}u_{i+1}\cdots u_n$ where $\overline{u_i}$ means the complement of $u_i$. By the definition of $n$-cube, two vertices have common neighbors iff their labels differ in exactly 2 bit positions. So any pair of vertices either have no common neighbors or have exactly 2 common neighbors. Thus the girth of an $n$-dimensional hypercube $g(Q_n)$ is 4 when $n\ge 2$.

In \cite{Caruso2002}, Caruso et.al. have determined the diagnosability of $n$-dimensional hypercube is $n$ when $n\ge 1$.
\begin{lem}\label{dqn}
For $n\ge 1$, the diagnosability of an $n$-cube $Q_n$ under the PMC model is $n$.
\end{lem}

\subsection{$h$-restricted diagnosability and $1$-restricted edge diagnosability of hypercubes }
\begin{thm}
Let $Q_n$ be an $n$-dimensional hypercube, $1\le h\le n$, then $t_h^{e}(Q_n)= n-h$.
\end{thm}
{\textbf Proof}  1) Let $u=0^n$, $F_1=\{ u^{h+1}, \cdots u^{n}\}$, $F_2=\{u, u^{h+1}, \cdots u^{n}\}$, $S_1=\{uu_1, uu_2, \cdots uu_h\}$ $S_2=\emptyset$, . It's obvious that $(F_1, S_1)$ and $(F_2, S_2)$ are both consistent faulty pair with $|\max(|S_1, S_2|)|=h$.
By Lemma \ref{dis}, $(F_1, S_1)$,  $(F_2, S_2)$ are indistinguishable.  Thus $t_h^e(Q_n)\le \max(|F_1|, |F_2|)|-1 =  n-h$.

2) In this paragraph we use contradiction to prove that  $t_h^e(Q_n)\ge n-h$. Suppose not, that is,  $t_h^e(Q_n)< n-h$. So there exist two indistinguishable consistent faulty pairs $(F_1, S_1)$ and $(F_2, S_2)$  with $\max(|S_1|, |S_2|)\le h$ and $\max(|F_1|, |F_2|)\le n-h$. Since$(F_1, S_1)$ and $(F_2, S_2)$ are two distinct  consistent faulty pairs,  either $F_1\Delta F_2\not = \emptyset$ or $S_1\Delta S_2\not = \emptyset$.

Case 1) $F_1\Delta F_2\not = \emptyset$. Then either  $F_1 -F_2$ or $F_2 -F_1$ has at least 2 vertices or $\max\{|F_1-F_2|, |F_2-F_1|\}=1$.

Subcase 1.1) $|F_1-F_2|=1, |F_2-F_1|=1$.

Suppose $F_1-F_2=\{u\}, F_2-F_1=\{v\}$. For any vertex $w$ in $N_{V-F_1\cup F_2}(u)$, $(u,w)$ must be in $S_2$  since $(F_1, S_1)$, $(F_2, S_2)$ are indistinguishable consistent pairs. Since $|S_2|\le h$, $u$ has at least $n-h$ vertices in $F_2$. Considering $|F_2|\le n-h$. We have $F_2\subset N(u)$ and $|F_2|=n-h$. Similarly, we can prove that $|F_1|=n-h$ and $F_1\subset N(v)$. Thus $u, v$ are adjacent and they have $n-h-1$ common neighbors. This is impossible since $g(Q_n)=4$.

Subcase 1.2) $|F_1-F_2|=1, |F_2-F_1|=0$ or  $|F_1-F_2|=0, |F_2-F_1|=1$.

Without loss of generality, suppose  $|F_1-F_2|=1, |F_2-F_1|=0$. Suppose $F_1-F_2=\{u\}$, then similarly as in the above analysis, we can prove that $|F_2|=n-h$, so $|F_1|=n-h+1$ contradicting with $|F_1|\le n-h$.

Subcase 1.3) $|F_1-F_2|\ge 2$ or  $|F_2-F_1|\}\ge 2$.

Without loss of generality, $|F_1-F_2|\ge 2$. Suppose $u, v\in F_1 -F_2$. For any vertex $w\in N_{V-F_1\cup F_2}(u, v)$, the edge between $w$ and $u$ or $v$ must be in $S_2$ since $(F_1, S_1)$ and $(F_2, S_2)$ are indistinguishable consistent pairs. So $|N_{F_1\cup F_2}(u, v)|= |N(u,v)| - |N_{V-F_1\cup F_2}(u, v)|\ge  2n-2 -h$. Thus $|F_1\cup F_2|\ge 2+(2n-2-h)=2n-h>|F_1|+|F_2|$, a contradiction.

Case 2) $S_1 \Delta  S_2\not = \emptyset$

Without loss of generality, suppose $S_1 -S_2\not =\emptyset$. Suppose $e=(u, v)\in S_1 -S_2$. Since $(F_1, S_1)$ and $(F_2, S_2)$ are indistinguishable consistent pairs, $u, v\not\in F_1$ and at least one of $u, v$ is in $F_2$. That is, $F_1\Delta F_2\not =\emptyset$, similar proof as in case 1) can obtain a contradiction.

\begin{thm}
$S_1(Q_n)=n-2$ for $n\ge 2$.
\end{thm}
Proof: 1) By Lemma \ref{diagboudary}
$S_1(Q_n)\le n-2$.

2) We use contradiction to prove that $S_1(Q_n)\ge n-2$. Suppose $S_1(Q_n)< n-2$, that is, there exists two indistinguishable consistent pairs $(F_1, S_1)$, $(F_2, S_2)$ with $|F_1|, |F_2|\le 1, |S_1|, |S_2|\le n-2$. Since $S_0(Q_n)=n*2^{n-1}$, at least one of $F_1, F_2$ is not empty. Without loss of generality, suppose $F_1=\{u\}$.

Subcase 2.1) $F_2=\emptyset$. In this case any edge in $NE(u)$ must be in $S_2$ since $(F_1, S_1)$, $(F_2, S_2)$ are indistinguishable. Thus $n-2\ge |S_2|\ge n$, a contradiction.

Subcase 2.2) $F_2 =F_1$. By lemma \ref{diagboudary}, this is impossible.

Subcase 2.3) $F_2=\{v\}$, any edge in $NE(u)$ other than $uv$ must be in $S_2$ since $(F_1, S_1)$, $(F_2, S_2)$ are indistinguishable. Thus $|S_2|\ge n-1$, contradicting to our assumption that $|S_2|\le n-2$. \hfill\rule{1mm}{2mm}
%
%
%
%
%

\section{Conclusions}
In this paper, we make two contributions: 1, To adapt to the hybrid faulty circumstances, we generalize the well-known PMC model to obtain the generalized PMC model. Based on a justified assumption that all incident edges of a faulty vertex is good, we propose the definition of consistent faulty pairs and characterize distinguishable consistent faulty pairs. Thus the foundation of generalized PMC model is established. 2. We then generalize diagnosability to get the $h$-edge restricted diagnosability and $r$-vertex-restricted diagnosability for measuring the fault  diagnosis capability of interconnection networks under hybrid fault circumstances. We then explore the $h$-edge-restricted diagnosability and $r$-vertex-restricted edge diagnosability of hypercubes. The result is important for understanding the fault diagnosis capability of hypercubes under hybrid fault circumstances. The method used here may be used to explore the two parameters of other interconnection networks under the GPMC model.

Here we give some advice for possible future work:

1. the $r$-vertex restricted diagnosability of hypercubes when $r\ge 2$ may be explored.

2. Similar methods used here may be used to explore the $h$-edge restricted diagnosability and $r$-vertex-restricted diagnosability  of other interconnection networks.

3. The GPMC model may be used to the fault diagnosis in wireless sensor networks.

\end{document}